\documentclass{cta-author}

\makeatletter
\def\@journal@title{Preprint}
\renewcommand{\jourinf}{}
\renewcommand{\oddpagefooter}{\hbox to \textwidth{\hfill\thepage}}
\renewcommand{\evenpagefooter}{\hbox to \textwidth{\hfill\thepage}}
\makeatother

{}
{}
{}
\usepackage{graphicx}
\usepackage{booktabs}
\usepackage{array}
\usepackage{amsmath}
\usepackage{hyperref}
\hypersetup{hidelinks}
\usepackage{tabularx}
\usepackage{algorithm}
\usepackage{algpseudocode}
\usepackage{mdframed}
\usepackage{multirow}
\usepackage{xcolor}
\usepackage{amssymb}
\usepackage{pifont}
\newcommand{\cmark}{\ding{51}}
\newcommand{\xmark}{\ding{55}}

\begin{document}

\title{Detect First, Explain Later: Training-Free Temporal-Memory Digital Twin Anomaly Detection with Post-Hoc LLM Interpretation for ICS}

\author{\au{Konstantinos E. Kampourakis$^{\corr1}$}, \au{Vasileios Gkioulos$^{1}$}, \au{Sokratis Katsikas$^{1}$}}

\address{\add{1}{Department of Information Security and Communication Technology, Norwegian University of Science and Technology (NTNU), Gjovik, 2802, Norway}
\email{konstantinos.kampourakis@ntnu.no}}

\begin{abstract}
Industrial Control Systems (ICS) are increasingly exposed to cyber-physical attacks that manifest as subtle and temporally evolving deviations in process behavior. Detecting such anomalies requires reasoning over persistence, cross-signal dependencies, and process-level constraints. Digital Twins (DTs) encode system knowledge through physical and logical relationships between signals, but existing DT-based approaches rely on instantaneous rule violations and lack mechanisms to aggregate weak evidence over time. This paper proposes a training-free anomaly detection method that combines deterministic DT constraints with explicit temporal memory. The DT monitors process signals and produces anomaly scores based on constraint violations, while a lightweight memory mechanism captures persistence and contextual relationships across time. The approach is evaluated on the HAI and BATADAL datasets. Ablation results show that the memory-less detector fails completely, demonstrating that temporal aggregation is essential for DT-based detection. On HAI, the memory-aware DT achieves stable detection with only 4 false alarm events, and on BATADAL, it remains effective without retraining, with 21 false alarms under domain shift. In comparison, Isolation Forest (IF) produces substantially more false alarms (328 on HAI and 127 on BATADAL), while Autoencoder (AE) exhibits dataset-dependent behavior, achieving high precision on BATADAL but low recall and inconsistent performance overall. A gated LLM is used for post-hoc interpretation, providing structured explanations without affecting detection performance. Our findings highlight the importance of temporal memory in constraint-based detection and support the use of decoupled reasoning for interpretability in ICS monitoring.
\end{abstract}

\maketitle

\section{Introduction}
\label{S:Intro}
Industrial Control Systems (ICS) are playing an ever more pivotal role in Industry 5.0, a paradigm that prioritizes human-centered, flexible, and highly interconnected industrial operations~\cite{alcaraz2024digital}. This shift introduces tighter integration between physical processes, digital systems, and operator interaction, but also expands the attack surface of critical infrastructure. In this setting, cyber-physical attacks do not necessarily cause immediate system failures, but often manifest as gradual and subtle deviations in process behavior.

Real-world incidents consistently illustrate this challenge. For instance, in the 2021 Oldsmar water treatment attack~\cite{cervini2022don}, an adversary gained access to an operator workstation and attempted to manipulate chemical dosing parameters. Similarly, the well-known Stuxnet incident~\cite{farwell2011stuxnet} demonstrated that malicious control logic can alter physical processes while masking its effects from operators. These cases show an important struggle in ICS monitoring: attacks are often observable only through small, temporally evolving inconsistencies in system behavior.

The most recent ICS anomaly detection approaches rely on data-driven models that learn patterns from historical data and attempt to generalize to new conditions~\cite{ayodeji2020new, raman2021hybrid}. While these may be effective in controlled settings, their models are often sensitive to distribution shifts, require retraining, and provide limited insight into the underlying causes of detected anomalies~\cite{bhatt2025enhancing,fung2024attributions}. In operational environments, the lack of interpretability is a significant limitation. Alerts must be understood in terms of process behavior, and therefore, an effective detection approach should remain interpretable, minimize reliance on retraining, and ground its decisions in the structure of the controlled process.

To that end, Digital Twin (DT) models provide a structured alternative, as they embed physical and logical knowledge of the system into the detection process~\cite{Kampourakis2025, homaei2024review}. Constraint-based DT approaches can identify anomalies as violations of process invariants and enable transparent and interpretable detection. However, current DT-based approaches typically depend on instantaneous rule evaluations and lack the capability to reliably accumulate weak or partial evidence over time~\cite{Kampourakis2026,alhamam2025comprehensive}. In ICS settings, detecting anomalies at the event level is often more operationally meaningful than identifying isolated point anomalies, since attacks progress over time and may manifest as subtle, yet persistent deviations~\cite{kampourakis2026systematic,lugaresi2023online}. Consequently, such methods can struggle to differentiate brief fluctuations from prolonged abnormal behavior.

In parallel, recent progress in Large Language Models (LLMs) indicates that they may facilitate post-hoc interpretation, provided that their outputs are anchored in physically meaningful evidence~\cite{krishna2023post}. These characteristics position LLMs as a promising means for explaining anomalies and generating higher-level summaries of system behavior. Nonetheless, naively deploying LLMs on ICS data can introduce concerns about reliability, hallucinations, and the necessity of grounding their reasoning in physically interpretable evidence~\cite{majeed2024reliability}.

This paper addresses these challenges by proposing a training-free anomaly detection approach for ICS based on DTs, temporal context, and optional post-hoc interpretation. The approach is designed to remain interpretable, reduce reliance on retraining, and provide consistent detection of temporally evolving anomalies in industrial processes. An additional interpretation layer is incorporated to support the understanding of detected events without affecting the detection process.

The main contributions of this work are:

\begin{itemize}
\item A training-free, constraint-driven DT-based anomaly detection method for ICS that operates without model training and provides interpretable detection based on process-level invariants.
\item A temporal memory mechanism that captures persistence and contextual relationships, shown to be essential for effective detection in ICS environments.
\item A gated post-hoc LLM interpretation layer that provides structured explanations without affecting detection decisions.
\item A cross-dataset evaluation on HAI~\cite{shin2020hai} and BATADAL~\cite{taormina2018battle}, demonstrating reliable detection on structured industrial processes and partial transferability under domain shift.
\item An empirical analysis of the trade-offs between precision, recall, detection delay, and false alarms in constraint-based detection.
\end{itemize}

The remainder of the paper is organized as follows. Section~\ref{S:Related} reviews related work on ICS anomaly detection, DT approaches, and LLM-based reasoning. Section~\ref{S:Method} presents the proposed method, including DT constraint modeling, temporal memory, and LLM integration. Section~\ref{S:Exper} describes the experimental setup and evaluation methodology. Section~\ref{S:Results} presents and analyzes the results. Section~\ref{S:Dis} discusses limitations and implications for deployment, and Section~\ref{S:Con} concludes the paper.

\section{Related Work}
\label{S:Related}

In this section, we review existing approaches to anomaly detection in ICS, with a focus on three main directions: data-driven methods, DT-based detection, and explainability and reasoning techniques. These directions differ in how they model system behavior, handle temporal dynamics, and provide interpretability. Data-driven approaches focus on learning patterns from historical data, DT-based methods rely on process-level knowledge and constraints, while explainability-oriented work seeks to make detection outcomes more interpretable through feature attribution or higher-level reasoning.

The work in~\cite{mbasso2026digital} presents a DT-enabled anomaly detection framework for industrial systems that combines data-driven temporal modeling with process-aware monitoring. The approach leverages LSTM-based sequence learning to capture system dynamics, Isolation Forest (IF) to estimate operating envelopes, and a fusion mechanism to combine multiple anomaly indicators. A DT is used to contextualize detections, support visualization, and enable operator interaction, including counterfactual analysis and triage. The framework emphasizes time-aware detection and event-level evaluation.

The authors in~\cite{kausar2025digital} combine DT modeling with XAI to improve detection accuracy and operator interpretability in industrial IoT settings. The method integrates real-time sensor data with simulation models and uses SHapley Additive exPlanations (SHAP), counterfactual explanations, and natural-language normalization to make anomaly decisions more actionable for engineers and operators. This line of work is important because it shows that DTs can support both detection and explanation, but the explanation layer remains based on classical XAI techniques.

The contribution in~\cite{birihanu2025explainable} focuses on temporal correlations among ICS devices and uses sliding windows, Pearson correlation, latent correlation matrices, and a multivariate Gaussian model to detect anomalies. The authors then apply SHAP to identify which correlated features contribute most to anomalous predictions, which gives the method a root-cause-oriented explanation layer. This paper emphasizes correlations, windowing, and explainability in ICS anomaly detection, but it remains a purely statistical and feature-driven approach, without incorporating DT-based constraints or higher-level reasoning mechanisms.

Sayghe et al.~\cite{sayghe2025digital} propose a DT-based IDS for SCADA that combines high-fidelity process simulation, live sensor modeling, adversarial attack injection, and hybrid anomaly detection. The framework fuses physics residuals with Machine Learning (ML), using an LSTM-based cyber module and a one-class SVM physical module, and is evaluated on a simulated water treatment plant under false data injection, DoS, and command-injection attacks. This work demonstrates how DTs can support process-aware intrusion detection in SCADA, focusing on hybrid ML-based detection without explicit temporal reasoning mechanisms.

The work in~\cite{alcaraz2024digital} proposes a DT plus ML framework for online protection in industrial environments, with emphasis on advanced and stealthy threats. The authors argue that the DT improves anomaly detection by comparing expected and observed behavior in a real industrial control testbed, and they report that the approach is effective against malicious perturbations in critical system sections. They use DTs as a defensive abstraction for discrepancy detection, focusing on model-based detection of process deviations.

The authors in~\cite{krishnaveni2024cyberdefender} propose a multi-layer DT-ICPS security architecture that combines honeynets, SDN, XAI-based feature selection, and a BO-GRU-LSTM intrusion detector. The framework is evaluated on SWaT, CICIDS-2018, and a honeypot dataset, and it emphasizes both detection performance and explainability through SHAP and ensemble-based filter feature selection. The main focus of this work is on supervised intrusion classification with integrated XAI components.

Varghese et al.~\cite{varghese2022digital} extend an open-source ICS DT with an ML-based IDS and multiple process-aware attack scenarios, including command injection, network DoS, calculated measurement modification, and naive measurement modification. The paper evaluates several supervised classifiers and then builds a stacked ensemble that achieves near real-time intrusion detection, while also showing that the original SIEM-style correlation module misses process-aware attacks beyond network DoS. Its main contribution is the use of DTs as a testbed for generating labeled attack data and benchmarking classifiers, primarily as a testbed for generating labeled data and evaluating supervised classifiers.

\begin{table*}[t]
\centering
\scriptsize
\setlength{\tabcolsep}{3pt}
\renewcommand{\arraystretch}{1.1}
\begin{tabularx}{\textwidth}{
p{1cm} 
>{\raggedright\arraybackslash}X 
c 
c 
>{\raggedright\arraybackslash}X 
>{\raggedright\arraybackslash}X}
\hline
\textbf{Year} & \textbf{Paper} & \textbf{DT} & \textbf{Explainability} & \textbf{Main Idea} & \textbf{Key Characteristics} \\
\hline

2026 & \cite{mbasso2026digital} & \cmark & Rule-based & DT-based anomaly detection with LSTM + IF fusion & LSTM + IF, fusion, DT interface \\

2025 & \cite{kausar2025digital} & \cmark & XAI & DT-based IIoT anomaly detection with XAI & SHAP, counterfactuals, DT + ML \\

2025 & \cite{birihanu2025explainable} & \xmark & XAI & Correlation-based ICS anomaly detection & Pearson correlation, Gaussian model, SHAP \\

2025 & \cite{sayghe2025digital} & \cmark & -- & DT-driven SCADA IDS with hybrid ML detection & LSTM + SVM, physics + ML \\

2024 & \cite{alcaraz2024digital} & \cmark & -- & DT-assisted anomaly detection for industrial systems & DT + ML, testbed validation \\

2024 & \cite{krishnaveni2024cyberdefender} & \cmark & XAI & DT-based ICPS security with deep learning IDS & BO-GRU-LSTM, SDN, honeynet, SHAP \\

2022 & \cite{varghese2022digital} & \cmark & -- & DT-based IDS with ML benchmarking & Supervised ML, DT testbed \\

\hline
2026 & This Work & \cmark & LLM (gated) & Training-free DT-based anomaly detection with memory & Constraint DT, temporal memory, LLM \\
\hline

\end{tabularx}
\caption{Comparison of DT-based and explainable anomaly detection approaches for ICS.}
\label{T:related_work}
\end{table*}

Table~\ref{T:related_work} summarizes representative DT-based and explainable anomaly detection approaches for ICS. Existing works primarily fall into two categories. The first group integrates DTs with ML models for process-aware monitoring, relying on learned representations of system behavior. The second group focuses on explainability, typically through XAI techniques such as SHAP or correlation-based analysis, which provide feature-level insights into detection outcomes. Across both directions, most approaches depend on training data and model calibration, which can limit robustness under changing system conditions. In addition, temporal aspects of anomalies are typically handled implicitly through learned models rather than explicit mechanisms for aggregating evidence over time. In contrast, the work at hand explores a constraint-based approach to anomaly detection that emphasizes temporal aggregation and interpretability. The proposed method explicitly separates detection from interpretation, enabling detection based on process-level constraints and temporal context, while supporting post-hoc reasoning without influencing detection decisions.

\section{Methodology}
\label{S:Method}

In this section, we describe the proposed methodology for training-free anomaly detection in ICS using DT constraints, temporal memory, and gated post-hoc LLM interpretation. The main objective of our method is to separate detection from reasoning, so that anomaly scoring remains deterministic while the LLM is used only to explain detected behavior.

\subsection{Overview}

The proposed method is a training-free anomaly detection system that combines constraint-based DT modeling with temporal memory and optional LLM-assisted interpretation. The DT continuously monitors multivariate ICS signals and identifies deviations through violations of process-level constraints. A memory mechanism captures temporal persistence and contextual relationships across signals, enabling aggregation of weak or intermittent violations into robust event-level detections. An LLM is optionally invoked as a post-detection reasoning module, without influencing the detection process.

The system operates in three stages. First, incoming signals are evaluated against a set of physical and logical constraints. Second, constraint violations are aggregated over time using a sliding window and stored in DT memory. Third, a gating mechanism determines whether the LLM should be invoked for interpretation. This design enforces a strict separation between detection and reasoning, ensuring that detection remains deterministic and grounded in system dynamics.

\subsection{Constraint-Based Detection and Temporal Modeling}

The DT represents the ICS as a set of interacting processes, such as flow, level, and actuation. At each time step $t$, a set of constraints $\mathcal{C} = \{c_1, c_2, \dots, c_n\}$ is evaluated over the system state $x_t$. Each constraint produces a violation score:
\begin{equation}
s_i(t) = f_i(x_t),
\end{equation}
where $f_i(\cdot)$ quantifies the deviation from expected physical or logical behavior.

The instantaneous DT score is defined as:
\begin{equation}
S_{\text{DT}}(t) = \max_{i} s_i(t),
\end{equation}
capturing the most severe violation at time $t$.

Isolated violations are often insufficient to indicate anomalies in ICS environments due to noise and transient fluctuations. To address this, the method incorporates temporal aggregation using a sliding window of length $W$:
\begin{equation}
S_{\text{window}}(t) = \max_{\tau \in [t-W, t]} S_{\text{DT}}(\tau).
\end{equation}

The sliding-window aggregation captures the maximum deviation within a temporal window, while persistence is modeled explicitly through the DT memory, which tracks sustained constraint violations and their temporal continuity. Event-level stability is further enforced through hysteresis-based post-processing.

In addition to window-based aggregation, the DT maintains a memory structure that explicitly models temporal persistence and contextual relationships. The memory stores:
\begin{itemize}
    \item active constraints and their severity levels
    \item persistence of violations through constraint streaks
    \item process domains involved (e.g., FLOW, LEVEL, ACTUATION)
    \item transitions and interactions across process domains
\end{itemize}

This memory allows the system to distinguish between transient deviations and sustained anomalous behavior. In practice, this mechanism is critical, as it enables the accumulation of weak evidence over time and supports stable event-level detection.

A detection decision is made using a threshold $\theta$:
\begin{equation}
\hat{y}(t) =
\begin{cases}
1, & \text{if } S_{\text{window}}(t) \geq \theta \\
0, & \text{otherwise}
\end{cases}
\end{equation}

The threshold $\theta$ defines the operating point of the detector and is selected to balance false alarms and missed detections. It does not correspond to a learned model parameter, but rather to a calibrated decision boundary over DT scores.

\subsubsection{Constraint Design}

The constraint set $\mathcal{C}$ encodes physical and logical relationships between ICS variables. Each constraint $c_i$ is implemented as a function $f_i(x_t)$ that quantifies deviation from expected system behavior. These functions include residual-based deviations, consistency checks, and threshold-based violation measures.

Constraints are derived from domain knowledge and fall into several categories: (i) flow–level consistency (e.g., inflow vs. tank level change), (ii) actuator–sensor consistency (e.g., valve state vs. flow response), (iii) operational bounds (e.g., valid sensor ranges), and (iv) rate-of-change constraints.

The proposed approach does require manual or semi-manual constraint engineering, particularly for process-aware relationships such as actuator–sensor consistency and flow–level dependencies. As the number of I/O points increases, the effort required to define and validate constraints also increases. However, the method is not intended to eliminate domain expertise. On the contrary, the aim is to leverage existing process knowledge, already available in engineering documentation, control logic, P\&IDs, alarm rules, and operational specifications.

Importantly, the manual definition of constraints does not scale linearly with the number of sensors or I/O points. Instead, the proposed approach follows a compositional modeling strategy in which process-level invariants are defined for standard industrial primitives, such as valve–tank systems, flow loops, pumps, and actuator–sensor relationships, and then instantiated across multiple subsystems.

As a result, the engineering effort scales primarily with the number of unique process types rather than the total number of monitored variables. In practice, many constraints can be derived directly from existing engineering artifacts, including P\&ID documentation, PLC logic, alarm rules, and process specifications. This allows the DT to leverage existing process knowledge while reducing the need for fully manual constraint definition in large-scale ICS environments.

Table~\ref{T:constraints} provides representative examples used in this study.

\begin{table}[h]
\centering
\caption{Representative DT constraints}
\label{T:constraints}
\begin{tabular}{l l}
\hline
Type & Description \\
\hline
Flow–Level & Inflow approximates change in tank level \\
Actuation & Valve open implies increase in flow \\
Range & Sensor within operational bounds \\
Rate & Change between time steps is bounded \\
\hline
\end{tabular}
\end{table}

\subsection{LLM-Assisted Interpretation and Gating}

The LLM is integrated as a post-detection reasoning component that operates on structured outputs produced by the DT. It does not modify detection decisions, but provides a semantic interpretation of observed system behavior. The LLM receives a structured input derived from DT memory, including indexed constraint violations (e.g., Constraint\_01, Constraint\_02), persistence indicators, affected process domains, and summaries of recent system behavior. Each constraint identifier corresponds to a specific DT rule or process relationship, enabling traceable interpretation and attribution. All inputs are derived exclusively from DT outputs to ensure grounding in system dynamics. The LLM used in this study is GPT-5-mini, accessed via API with a fixed prompt template. The model is accessed through a fixed prompt template designed to preserve consistent and grounded reasoning behavior across invocations.

The LLM output is constrained to a structured format that references the relevant constraint identifiers used during reasoning. This allows interpretations to remain grounded in DT-derived evidence and enables operator verification of the reasoning chain. This design also reduces the likelihood of unsupported interpretations by constraining the LLM to reason only over explicitly referenced DT evidence.

Also, to ensure controlled and reliable operation, the LLM is invoked through a gating mechanism. The gating decision is based on the DT score:
\begin{equation}
g(t) =
\begin{cases}
1, & \text{if } S_{\text{window}}(t) \geq \theta_g \\
0, & \text{otherwise}
\end{cases}
\end{equation}
where $\theta_g \leq \theta$.

This design allows the LLM to be triggered only when DT evidence exceeds a calibrated gating threshold. Importantly, all LLM inputs are derived exclusively from DT outputs, ensuring that reasoning remains grounded in physically meaningful evidence and reducing the risk of unsupported conclusions.

\subsection{Operational Workflow}

At each time step, the system performs the following operations:
\begin{enumerate}
    \item Evaluate DT constraints and compute $S_{\text{DT}}(t)$
    \item Update sliding window and DT memory
    \item Compute $S_{\text{window}}(t)$ and detection decision $\hat{y}(t)$
    \item Apply gating function $g(t)$
    \item If $g(t) = 1$, invoke the LLM for interpretation
\end{enumerate}

The key dependency is from constraints $\rightarrow$ memory $\rightarrow$ interpretation, not from the LLM back to detection.

\begin{figure}
    \includegraphics[width=\linewidth]{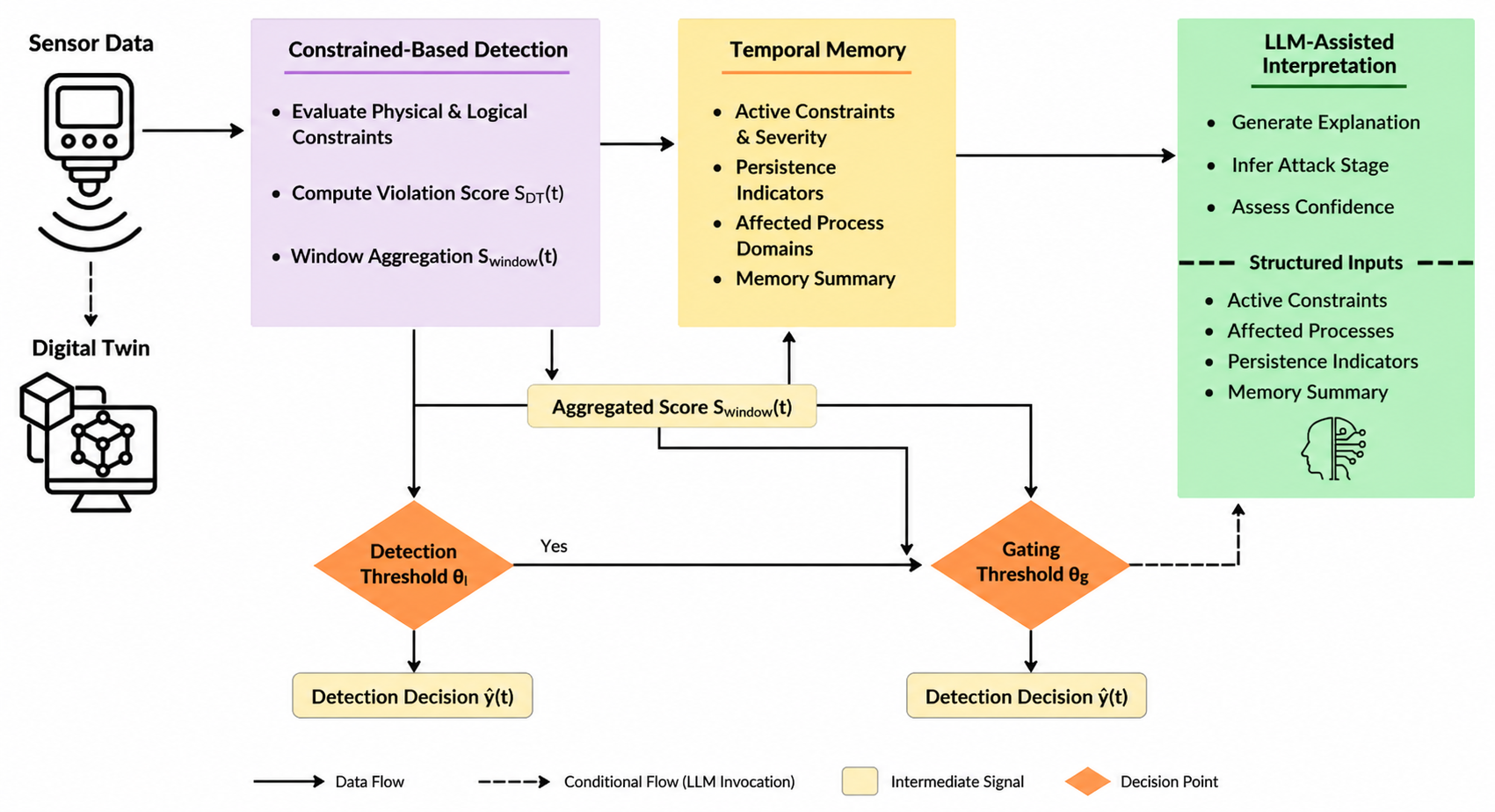}
    \caption{Methodology Overview}
    \label{F:Methodology}
\end{figure}

Figure~\ref{F:Methodology} provides an overview of the proposed method, showing the flow from DT-based constraint evaluation and temporal aggregation to detection and optional LLM-based interpretation.

The pseudocode in~\ref{A:dt} summarizes the proposed detection procedure. At each time step, the DT evaluates a set of physical and logical constraints over the observed system state and computes an instantaneous violation score. These scores are aggregated over a sliding window to obtain a robust signal that reduces sensitivity to transient fluctuations. In parallel, a temporal memory structure is updated to capture persistence of constraint violations and contextual relationships across process domains. This memory enables the accumulation of weak but consistent evidence over time, which is essential for reliable detection. A detection decision is then obtained by applying a threshold to the aggregated score. Independently, a gating condition determines whether the LLM should be invoked. When activated, the LLM operates on structured inputs derived exclusively from DT memory and produces a semantic interpretation of the detected anomaly. Importantly, the LLM does not influence the detection process, ensuring that detection remains deterministic and grounded in system dynamics.

\begin{algorithm}[H]
\caption{DT-based anomaly detection with temporal memory and gated LLM interpretation}
\label{A:dt}
\begin{algorithmic}[1]

\State Initialize DT memory $\mathcal{M}$
\State Initialize sliding window $\mathcal{W}$

\For{each time step $t$}
    \State Observe system state $x_t$
    
    \State Evaluate constraints $\mathcal{C}$ on $x_t$
    \State Compute violation scores $\{s_i(t)\}$
    
    \State $S_{\text{DT}}(t) \gets \max_i s_i(t)$
    
    \State Update window $\mathcal{W}$ with $S_{\text{DT}}(t)$
    \State Compute $S_{\text{window}}(t)$
    
    \State Update memory $\mathcal{M}$ with:
        \begin{itemize}
            \item active constraints
            \item persistence (streaks)
            \item process domains
        \end{itemize}
    
    \If{$S_{\text{window}}(t) \geq \theta$}
        \State $\hat{y}(t) \gets 1$ 
    \Else
        \State $\hat{y}(t) \gets 0$
    \EndIf
    
    \If{$S_{\text{window}}(t) \geq \theta_g$}
        \State Construct indexed DT evidence summary from $\mathcal{M}$
        \State Call LLM for interpretation
    \EndIf

\EndFor

\end{algorithmic}
\end{algorithm}

\section{Experimental Setup}
\label{S:Exper}

The experimental setup, described in this section, evaluates the proposed DT-based method under both structured and heterogeneous ICS conditions.

\subsection{Datasets}

Our method is evaluated on two ICS datasets with distinct characteristics, HAI and BATADAL. The dual dataset evaluation is an efficient way to test the method under different process dynamics and anomaly profiles.

Specifically, the HAI dataset consists of multivariate time-series measurements collected from an industrial control environment with multiple interconnected processes. It includes labeled attack scenarios that exhibit structured and temporally consistent behavior, where anomalies manifest as persistent deviations within specific subsystems.

The BATADAL dataset, on the other hand, represents a water distribution system with multivariate sensor and actuator data across interconnected hydraulic processes. Compared to HAI, BATADAL exhibits higher variability and noisier dynamics, with anomalies that are less localized and more difficult to distinguish from normal operation. Labels are provided at the point level, with binary indicators of attack presence. The two datasets differ substantially in process structure, noise, and anomaly expression, which is a major challenge when it comes to cross-dataset generalization.

These datasets provide complementary evaluation conditions, allowing assessment of both structured industrial processes and more heterogeneous, less constrained environments.

\subsection{Preprocessing and Signal Configuration}

Raw signals are processed using a sliding window approach. Each window contains $W = 30$ time steps with a step size of 1. No feature engineering is applied beyond basic preprocessing, as the DT operates directly on raw signals through constraint evaluation.

For BATADAL, the provided label column is directly mapped to the binary anomaly target. For HAI, labels are aligned with the time-series data and propagated to the window level, where a window is considered anomalous if any contained time step corresponds to an attack.

All signals are used as-is without dataset-specific feature selection, ensuring that the method operates consistently across datasets.

\subsection{Detection Configuration}

At each time step, constraint violations are evaluated and aggregated into a DT score. Temporal aggregation is performed using a sliding window, producing a window-level score $S_{\text{window}}(t)$.

Detection is performed using a threshold $\theta$ applied to the aggregated score. The detector is training-free, while the operating point is defined through dataset-specific threshold calibration to control false alarm rates. Due to differences in score distributions, the resulting operating point is dataset-dependent: 

\begin{itemize}
    \item HAI: $\theta \approx 0.60$
    \item BATADAL: $\theta \approx 0.25$
\end{itemize}

Thresholds are calibrated using only normal data from the training portion of each dataset, without access to attack labels. The calibration targets a predefined false positive rate, ensuring that no test data is used in the threshold selection process. As such, the detector remains training-free, while the operating point is defined through dataset-specific calibration. The calibration step defines an operational decision boundary rather than learning detector parameters. Similar to threshold tuning in traditional IDS deployments, it controls the operating point without modifying the underlying DT detection mechanism.

This reflects differences in system dynamics and constraint activation patterns across datasets. Because thresholds are fixed rather than learned, performance is sensitive to dataset dynamics and operating conditions.

To improve event-level stability, a hysteresis-based post-processing mechanism is applied. Detection events are triggered when the score exceeds an entry threshold and remain active until the score falls below a lower exit threshold. Short gaps between detections are merged, and events below a minimum duration are filtered. This reduces fragmentation and suppresses spurious detections.

\subsection{LLM Configuration}

The LLM is used as an auxiliary reasoning module and does not influence detection decisions. It is invoked selectively based on a gating threshold $\theta_g = 0.6$, allowing interpretation during both confirmed and emerging anomalies.

The LLM input consists of structured summaries derived from DT memory, including active constraints, affected processes, and temporal persistence indicators. The output includes a textual explanation, an inferred attack stage when applicable, and a confidence estimate.

To control computational cost and avoid redundant queries, a cooldown mechanism enforces a minimum interval of 60 time steps between consecutive LLM calls, with a maximum of 20 calls per file.

No task-specific fine-tuning or prompt optimization is performed, ensuring consistent and reproducible behavior across runs. The average response time per invocation was approximately 1-2 seconds, depending on input length and system load. Since LLM calls are gated and infrequent, their impact on overall system latency remains limited.

Table~\ref{T:params} summarizes the key parameters used across all experiments. Ranges reflect minor adjustments across datasets.

\begin{table}[h]
\centering
\caption{Experimental parameters}
\label{T:params}
\begin{tabular}{l c}
\hline
\textbf{Parameter} & \textbf{Value} \\
\hline
Window size $W$ & 30 \\
Step size & 1 \\
Target train FPR & 0.005 \\
Threshold $\theta$ (HAI) & 0.60 \\
Threshold $\theta$ (BATADAL) & 0.25 \\
Gating threshold $\theta_g$ & 0.60 \\
Exit threshold fraction & 0.75--0.88 \\
Merge gap (steps) & 6--10 \\
Minimum event length & 5--8 \\
LLM cooldown & 60 steps \\
Max LLM calls & 20 per file \\
\hline
\end{tabular}
\end{table}

\subsection{Evaluation Metrics}

Performance is evaluated at both the window and event levels.

Window-level metrics include precision, recall, and F1-score, computed over all time steps. These metrics capture the point-wise classification performance of the detector.

Event-level evaluation focuses on the detection of attack occurrences. Consecutive anomalous windows are grouped into events, and a detected event is considered correct if it overlaps with a true attack interval. Event recall measures the fraction of true attack events that are detected, while false alarm events correspond to predicted events with no overlap.

Detection delay is defined as the average number of time steps between the start of a true attack event and the first corresponding detection. This metric reflects the responsiveness of the system under persistence-based detection.

The contribution of the LLM is evaluated qualitatively by analyzing its ability to provide consistent, grounded interpretations of detected anomalies and to abstain under insufficient evidence.

\subsection{Implementation Details}

All experiments are conducted in an offline setting. The method is implemented in Python using a 128GB RAM Windows 11 machine, with deterministic DT constraint evaluation and optional LLM calls executed through an external interface. Experiments are performed independently on each dataset using the same method and parameter structure. Differences in performance arise from dataset characteristics and threshold calibration, rather than structural changes to the method. Results are recorded in tabular form and analyzed at both the aggregate and per-file level, enabling detailed examination of detection behavior, temporal dynamics, and LLM interaction. The evaluation is offline and benchmark-based, which limits claims about real-time deployment.

\begin{figure}
    \includegraphics[width=\linewidth]{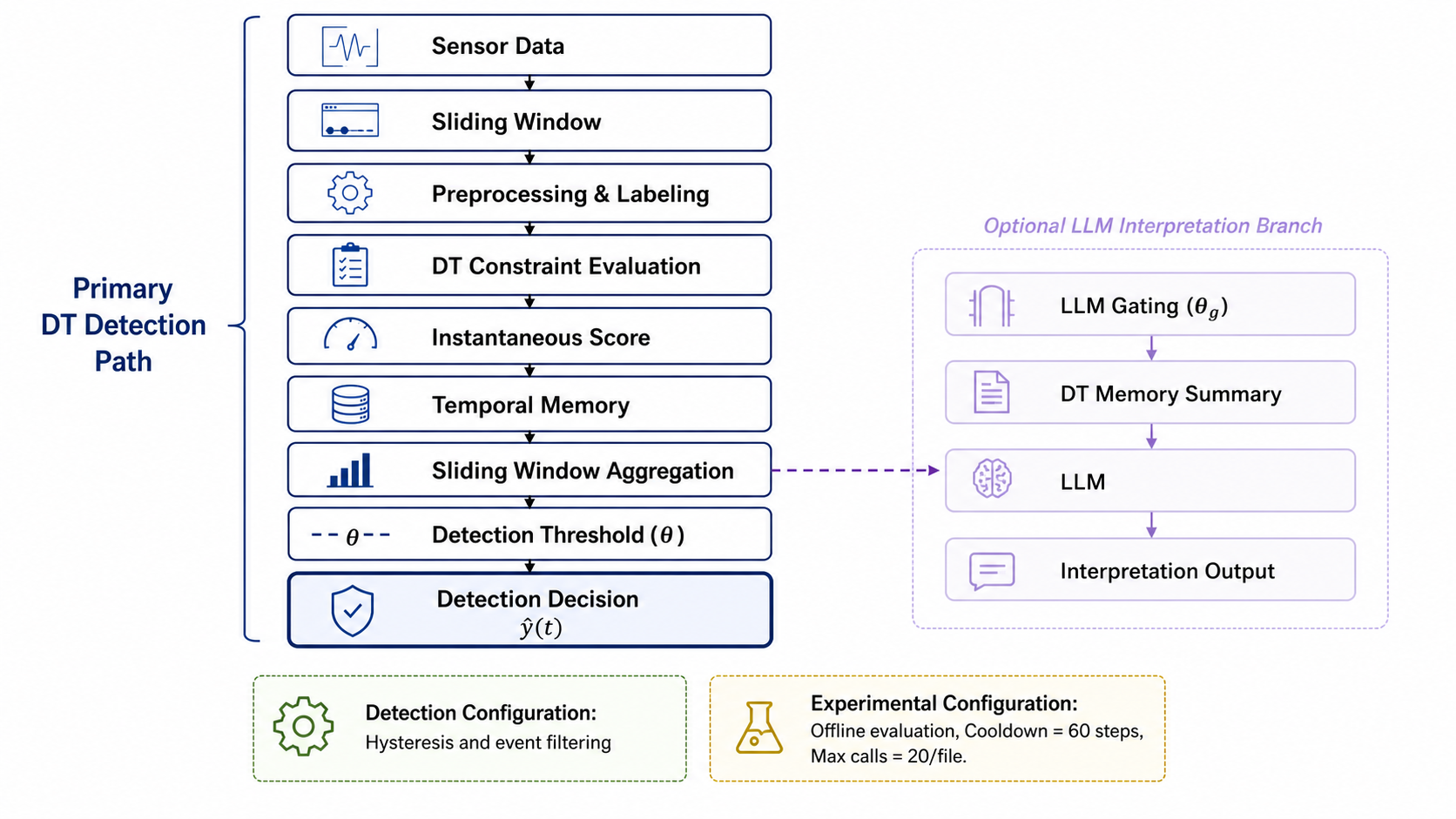}
    \caption{Overview of the proposed DT-based anomaly detection workflow.}
    \label{F:Exper}
\end{figure}

The overall experimental procedure is illustrated in Figure~\ref{F:Exper}. 

\section{Results}
\label{S:Results}

This section presents a comprehensive evaluation of the proposed DT-based anomaly detection method across HAI and BATADAL. The analysis focuses on three aspects: (i) the role of temporal memory, (ii) the contribution of LLM-assisted reasoning, and (iii) the robustness of the approach under cross-dataset deployment.

\subsection{Ablation Analysis and Role of Temporal Memory}

We begin by evaluating the contribution of temporal memory and LLM integration through an ablation study. Three configurations are considered: a purely instantaneous DT detector without memory, the proposed DT detector with temporal memory, and the full system augmented with LLM-based reasoning.

Table~\ref{T:hai_ablation} reports aggregated results on the HAI dataset. The results reveal a clear and decisive pattern. The DT detector without temporal memory fails completely, yielding zero true positives and zero detected events across all files. This result is consistent across both datasets and indicates that instantaneous constraint evaluation alone is insufficient for anomaly detection in ICS environments. Detection capability emerges only when temporal persistence is incorporated.

Introducing temporal memory fundamentally alters the detector's behavior. With memory enabled, the system achieves a micro-averaged F1 score of 0.31 and an event recall of 0.655. This represents a complete transition from non-functional to operational detection. The improvement reflects the importance of temporal persistence for DT-based anomaly detection.

The LLM does not measurably affect detection performance, confirming that it functions as an interpretation layer rather than a detector. As shown in Table~\ref{T:hai_ablation}, the DT+memory and DT+memory+LLM configurations produce nearly identical results across all metrics. This indicates that the LLM does not contribute to detection accuracy, but instead serves as an auxiliary reasoning mechanism. This negative result is important, as it isolates the source of detection capability within the deterministic DT and temporal memory components. The memory-less detector fails completely, so the ablation isolates temporal persistence as the enabling mechanism.

\begin{table}[h]
\centering
\caption{Ablation results on HAI dataset (aggregated)}
\label{T:hai_ablation}
\begin{tabular}{lcccc}
\hline
Configuration & Precision & Recall & F1 & Event Recall \\
\hline
DT (no memory) & 0.000 & 0.000 & 0.000 & 0.000 \\
DT + memory & 0.237 & 0.449 & 0.310 & 0.655 \\
DT + memory + LLM & 0.236 & 0.447 & 0.309 & 0.655 \\
\hline
\end{tabular}
\end{table}

A similar pattern is observed on the BATADAL dataset, as shown in Table~\ref{T:batadal_ablation}. The memory-less detector again fails, reinforcing the conclusion that temporal reasoning is indispensable. With memory enabled, the detector achieves non-zero detection performance, although at lower levels compared to HAI. The LLM again provides negligible benefit.

\begin{table}[h]
\centering
\caption{Ablation results on BATADAL dataset}
\label{T:batadal_ablation}
\begin{tabular}{lcccc}
\hline
Configuration & Precision & Recall & F1 & Event Recall \\
\hline
DT (no memory) & 0.000 & 0.000 & 0.000 & 0.000 \\
DT + memory & 0.129 & 0.247 & 0.170 & 0.600 \\
DT + memory + LLM & 0.129 & 0.247 & 0.170 & 0.600 \\
\hline
\end{tabular}
\end{table}

Taken together, these results show that temporal memory is the primary factor enabling detection, while the LLM has no measurable impact on detection performance.

\subsection{Detection Performance on HAI}

Afterwards, we examine the full system performance on the HAI dataset. Table~\ref{T:hai_results} reports per-file results for the DT+memory+LLM configuration. The most important observation is the combination of moderate precision, low false alarm rates, and stable event-level detection across files.
\begin{table*}[h]
\centering
\caption{Per-file detection results on HAI dataset}
\label{T:hai_results}
\begin{tabular}{lcccccc}
\hline
File & Precision & Recall & F1 & Event Recall & False Alarm Events & Delay \\
\hline
1 & 0.852 & 0.475 & 0.610 & 0.60 & 0 & 33 \\
2 & 0.526 & 0.197 & 0.286 & 0.35 & 5 & 122 \\
3 & 0.339 & 0.206 & 0.256 & 0.50 & 9 & 149 \\
4 & 0.776 & 0.629 & 0.695 & 0.80 & 2 & 37 \\
5 & 0.671 & 0.486 & 0.564 & 0.58 & 4 & 45 \\
\hline
Mean & 0.633 & 0.399 & 0.482 & 0.566 & 4.0 & 77.2 \\
\hline
\end{tabular}
\end{table*}

Several important observations emerge. First, the detector achieves moderate precision while maintaining low false alarm behavior across files, reaching a precision of 0.85 in some cases. This indicates that the system produces a low rate of false alarms, a critical requirement in ICS environments where excessive alarms are operationally unacceptable. The number of false alarm events remains low across all files, further reinforcing this property.

Second, event-level detection is robust. Event recall ranges from 0.35 to 0.80, with a mean value of 0.566. This demonstrates that the detector is capable of capturing the majority of attack events, even when point-wise recall is moderate.

Third, detection delay varies across files, with higher delays observed in more complex scenarios. This behavior is directly linked to the persistence requirements embedded in the temporal memory. The detector prioritizes reliability over immediacy, requiring sustained evidence before triggering an alarm. While this increases delay, it significantly reduces spurious detections. Detection delay increases because the method intentionally waits for sustained evidence before firing.

Overall, the HAI results show that the proposed DT-based approach achieves a favorable balance between precision and event-level recall, without relying on any training data. The results are particularly notable given the fully deterministic nature of our method.

\subsection{Cross-Dataset Evaluation on BATADAL}

To assess generalization, we evaluate the same detector on the BATADAL dataset without any modification of constraints or retraining. The results are summarized in Table~\ref{T:batadal_results}.

\begin{table}[h]
\centering
\caption{Detection results on BATADAL dataset (threshold = 0.25)}
\label{T:batadal_results}
\begin{tabular}{lcccccc}
\hline
Precision & Recall & F1 & Event Recall & False Alarm Events & Delay \\
\hline
0.129 & 0.247 & 0.170 & 0.60 & 21 & 30 \\
\hline
\end{tabular}
\end{table}

The results indicate that the detector remains functional under cross-dataset deployment, achieving an event recall of 0.60 without any adaptation. However, performance is significantly lower than on HAI, particularly in terms of precision and false alarm events. The drop in precision indicates that the HAI-derived constraints do not fully capture BATADAL's dynamics.

This behavior highlights an important limitation of constraint-based DT methods: detection performance depends strongly on the alignment between the constraint set and the underlying process structure. While HAI contains tightly coupled multi-process signals with relatively consistent temporal behavior, BATADAL represents a hydraulic system with different temporal and structural characteristics. As a result, constraint violations become weaker and less discriminative, leading to increased false positives and reduced sensitivity.

The low point-wise F1 score also indicates that the detector does not precisely capture attack boundaries. Instead, the persistence-based design emphasizes sustained anomalous behavior, often detecting the central portions of attack intervals more reliably than their onset or recovery phases. This behavior is consistent with the event-oriented objective of the proposed method, which prioritizes stable event-level detection over highly reactive point-wise classification. 

Importantly, the detector does not collapse under transfer. Although the constraint alignment degrades across domains, the temporal aggregation and memory mechanisms continue to provide stable event-level behavior. This suggests that the observed cross-dataset robustness originates more from persistence-based reasoning than from direct transferability of the constraint definitions themselves. In large-scale deployments, this limitation could be mitigated through compositional modeling approaches in which reusable constraint templates are instantiated across repeated industrial process primitives.

To illustrate the role of the LLM as a post-detection reasoning component, we present an example interpretation generated during an attack interval in the HAI dataset.
\paragraph{Example LLM-Assisted Interpretation:}

At a representative time step during an attack interval in the HAI dataset, the DT identified persistent violations in flow and level constraints within a single process unit. The violations persisted across multiple windows and reflected a consistent mismatch between process measurements and actuator behavior.

Based on the structured summary derived from DT memory, the LLM produced the following interpretation:

\begin{mdframed}[
  backgroundcolor=gray!10,
  linecolor=black,
  linewidth=0.4pt,
  nobreak=true
]
\textbf{\textit{Verdict:}} Sustained inconsistency observed between flow measurements and tank level dynamics [Constraint\_04], combined with actuator-state mismatch [Constraint\_07]. The persistence and consistency of these violations suggest a potential integrity attack affecting process control.
\end{mdframed}

Across all files, the LLM was invoked selectively through the gating mechanism, resulting in a limited number of calls per file. The generated interpretations remained generally aligned with the DT-derived structured inputs and explicit constraint references. Under weak or ambiguous conditions, the LLM typically produced lower-confidence outputs or avoided strong conclusions. In cases where multiple constraint violations were simultaneously active, the fixed prompt structure instructed the model to prioritize explaining the most persistent and severe DT evidence rather than attempting to resolve conflicting physical interpretations. LLM response time per invocation was approximately 1--2 seconds, and due to infrequent usage, its impact on overall system latency was negligible. The example reflects the general structure and reasoning style observed across LLM-assisted interpretations.

\subsection{Comparison with Baseline Methods}

To provide context for the proposed approach, we compare the DT-based detector with an unsupervised IF baseline and an MLPRegressor Autoencoder (AE), on both HAI and BATADAL datasets. IF is trained on normal data with a contamination parameter set to match the expected anomaly rate. The AE is trained using reconstruction loss on normal samples, with anomaly scores derived from reconstruction error. Both baselines use the same sliding window representation and are evaluated under the same event-level post-processing as the DT-based method to ensure fair comparison.

On the HAI dataset, both baseline methods achieve higher point-wise recall (above 0.8 in most files), but at the cost of extremely low precision and a large number of false alarm events. Across all files, IF and AE produce more than 300 and 700 false alarm events, respectively, indicating highly fragmented and noisy detections.

In contrast, the DT-based detector achieves lower recall but consistently reduces false alarms to a mean value of 4 across files. While point-wise F1 remains modest, event-level detection remains consistent. This suggests that the DT approach better captures sustained process deviations, while the baseline methods are overly sensitive to transient fluctuations.

On the BATADAL dataset, the DT-based detector achieves moderate event-level detection performance while maintaining a relatively low number of false alarm events (21).

In comparison, IF achieves a higher recall, but generates a large number of false alarm events (127). AE, on the other hand, attains high precision (0.875) with very few false alarms, but at the cost of low recall, indicating that many attack events are not detected. While all methods detect at least part of the attack activity, the baselines tend to produce either fragmented detections (IF) or overly conservative behavior (AE), limiting their effectiveness at the event level.

These results highlight a trade-off between data-driven sensitivity and constraint-based stability. IF tends to maximize recall by reacting to local deviations, which results in a high number of false positives and fragmented detections. AE adopts a more conservative regime, reducing false alarms but missing a substantial portion of anomalies. In contrast, the DT-based approach leverages temporal aggregation and process constraints to produce more coherent event-level detections.

Overall, the DT-based method sacrifices some sensitivity in structured environments such as HAI, but provides more stable and interpretable detections with fewer false alarms. In more heterogeneous settings such as BATADAL, it maintains balanced detection performance while avoiding excessive false alarms, indicating partial transferability under domain shift.

Table~\ref{T:dt_vs_baselines} summarizes the comparison between the DT-based detector and the baseline methods across both datasets.

\begin{table}[t]
\centering
\caption{Comparison with baseline methods}
\label{T:dt_vs_baselines}
\scriptsize
\setlength{\tabcolsep}{3pt}
\renewcommand{\arraystretch}{1.0}
\begin{tabular}{l l c c c c c}
\hline
Dataset & Method & Prec. & Rec. & F1 & Ev. Rec. & FA \\
\hline

\multirow{3}{*}{HAI} 
& DT & 0.633 & 0.399 & 0.482 & 0.566 & 4 \\
& IF & 0.044 & 0.916 & 0.084 & 0.980 & 328 \\
& AE & 0.056 & 0.813 & 0.104 & 1.000 & 723 \\
\hline

\multirow{3}{*}{BAT} 
& DT & 0.129 & 0.247 & 0.170 & 0.600 & 21 \\
& IF & 0.229 & 0.743 & 0.350 & 1.000 & 127 \\
& AE & 0.875 & 0.112 & 0.199 & 0.600 & 2 \\
\hline

\end{tabular}
\end{table}

\section{Discussion}
\label{S:Dis}

In this section, we discuss the implications of the experimental results along four dimensions. First, we analyze the separation between detection and interpretation in ICS monitoring, emphasizing the roles of constraint-based DT modeling and LLM-assisted reasoning. Second, we examine the robustness of the proposed method under varying data conditions, highlighting differences between structured and heterogeneous environments. Third, we consider practical deployment aspects, including interpretability, training-free operation, and computational constraints. Finally, we outline the main limitations of the approach and identify directions for future work.

\subsection{Detection vs Interpretation in ICS Monitoring}

The results highlight a clear separation between anomaly detection and attack interpretation in ICS environments. The constraint-based DT component provides consistent detection by grounding decisions in physical and logical system behavior. Detection is driven entirely by constraint violations and their temporal aggregation, without reliance on learned models or statistical assumptions.

A key finding of this work is that temporal memory is essential for effective DT-based detection. Without memory, the detector fails to identify any attack events across both datasets. With memory, the system transitions to an operational regime, indicating that persistence is necessary to distinguish transient fluctuations from sustained anomalies.

The integration of an LLM introduces a complementary interpretation layer. However, experimental results confirm that the LLM does not influence detection, as it is excluded from the scoring path. Its role is therefore limited to post-hoc reasoning, translating constraint violations into structured interpretations and high-level attack-stage interpretations. Because the LLM operates only on structured DT-derived summaries with explicit constraint references, the interpretation process remains traceable and grounded in physically meaningful evidence. This separation of roles is important, as it preserves the determinism and reliability of DT-based detection while enabling additional interpretability without affecting core decisions.

Importantly, the absence of measurable detection improvements should not be interpreted as a limitation of the LLM component itself. The objective of the LLM is not to improve classification performance, but to support human understanding and operational interpretation while preserving deterministic detection behavior.

These findings suggest that, in ICS monitoring, detection and reasoning should be explicitly decoupled. Detection should remain grounded in system dynamics, while higher-level interpretation can be layered on top without interfering with the detection mechanism.

\subsection{Robustness Under Varying Data Conditions}

The cross-dataset evaluation reveals that the structure and clarity of process-level deviations strongly influence DT-based detection. In the HAI dataset, anomalies produce consistent and persistent constraint violations, enabling more stable detection with higher precision and stable event-level performance.

In contrast, the BATADAL dataset exhibits higher variability and less clearly defined anomaly patterns. As a result, constraint violations are weaker and less consistent, leading to reduced precision and increased false alarm events. Despite this degradation, the detector remains functional without retraining, achieving moderate event-level recall. This indicates that the DT abstraction retains partial validity across domains, even when constraint alignment is suboptimal.

The observed detection delays reflect the persistence-based nature of the proposed method. Because anomaly decisions are generated through temporal aggregation and hysteresis, the detector intentionally waits for sustained evidence before triggering an alert. This design reduces fragmented detections and suppresses false alarms, but introduces latency in attack identification.

The suitability of this trade-off depends on the operational context. In slower industrial processes or supervisory monitoring scenarios, delayed but stable alerts may be preferable to highly reactive systems with excessive false positives~\cite{adepoju2022advancing}. In contrast, safety-critical or fast-control environments may require lower-latency mechanisms operating alongside the proposed approach. The method is therefore better positioned as a stable first-line monitoring and interpretation layer rather than an ultra-fast protective control mechanism.

The behavior of the LLM under these conditions is consistent with its design. When provided with strong DT evidence, it produces coherent and structured explanations. Under weaker or ambiguous conditions, it tends to produce low-confidence outputs or abstain from strong conclusions. However, its effectiveness is bounded by the quality of DT-based detection, as it operates only on DT-derived signals.

\subsection{Implications for Practical Deployment}

The proposed method offers several practical advantages for ICS monitoring. The DT component provides transparent and interpretable detection based on physically meaningful constraints, supporting operator trust and facilitating diagnosis. In ICS environments, both false alarms and missed detections carry significant risks. Missed attacks can have severe consequences in safety-critical systems, while excessive false alarms can overwhelm operators and reduce trust in monitoring systems~\cite{koay2023machine}. The proposed method explicitly addresses this trade-off by prioritizing stable detection and low false alarm rates, while maintaining sufficient event-level coverage. 

The absence of training requirements simplifies deployment and avoids issues related to data availability and model generalization. In contrast to data-driven approaches that optimize for high benchmark accuracy through extensive training, the proposed method operates without any training. It uses predefined constraints and temporal aggregation, which allows for immediate deployment. As such, it can serve as a first-line monitoring mechanism, providing lightweight and interpretable anomaly monitoring before more complex or data-driven systems are deployed.

The LLM component can be integrated as an optional interpretability layer, providing structured explanations without influencing detection decisions. The gating mechanism ensures controlled invocation, limiting computational overhead and preventing unnecessary reasoning in stable conditions.

At the same time, system performance depends on the alignment between the constraint set and the underlying process dynamics. In environments where constraints accurately capture system behavior, detection is reliable and precise. In more heterogeneous systems, additional effort is required to design constraints that reflect domain-specific characteristics.

Overall, the proposed approach targets a different operating regime, prioritizing interpretability, stability, and low false alarm rates over maximizing point-wise performance, with minimal deployment overhead. In practical ICS settings, stable event-level behavior and manageable false alarm rates may be operationally preferable to highly reactive detectors that maximize point-wise recall at the cost of excessive alert fragmentation~\cite{fung2022perspectives}.

\subsection{Limitations and Future Directions}

Several limitations highlight directions for future work. The current constraint set focuses primarily on local process relationships, which may limit the ability of the DT to capture complex interactions across subsystems. Extending the DT to model richer interdependencies and cross-process dynamics could improve detection performance in large-scale or distributed ICS environments.

The approach also depends on the quality and completeness of the constraint design. Since constraints define the expected system behavior, inaccurate, overly simplistic, or incomplete constraints can lead to degraded detection performance, particularly under domain shift, as observed in the BATADAL evaluation. This places importance on domain knowledge and careful constraint engineering during deployment.

Threshold selection and gating are currently based on fixed parameters calibrated per dataset. While this provides explicit control over false alarm rates, it introduces sensitivity to operating conditions and system dynamics. In environments with varying noise levels or changing process behavior, fixed thresholds may not generalize well. Adaptive or data-driven thresholding strategies could improve robustness across different operating regimes.

The evaluation of LLM-based interpretation remains qualitative. While the results indicate that the LLM produces generally consistent and DT-grounded interpretations, the absence of annotated datasets with attack stages or progression patterns limits the ability to quantitatively assess reasoning quality. Developing such benchmarks would enable more systematic evaluation and comparison of interpretation capabilities. Future work could incorporate automated consistency verification between generated interpretations and referenced DT constraints to further improve interpretability assurance. 

Additionally, the evaluation is conducted in an offline setting. Although the method is lightweight and the LLM is gated to limit overhead, further work is required to assess real-time deployment aspects, including end-to-end latency, scalability under continuous monitoring, and integration with operational ICS systems. In addition, the method does not aim to match the point-wise accuracy of fully trained data-driven models, but instead prioritizes stability and low false alarm rates, which are often more critical in practice.

Overall, the results indicate that DT-based detection with temporal memory provides a lightweight and interpretable alternative to data-driven approaches, while LLM-based reasoning enhances interpretability without affecting detection decisions.

\section{Conclusion}
\label{S:Con}

This paper presented a training-free anomaly detection approach for ICS based on constraint-driven DT modeling, temporal memory, and optional LLM-assisted interpretation. Detection is performed using process-level constraints and temporal aggregation, while the LLM operates as a gated post-hoc component that provides structured interpretations without influencing detection decisions.

Experimental results on the HAI and BATADAL datasets show that temporal memory is essential for DT-based detection. Without memory, the detector fails to identify attack events, while the inclusion of memory enables consistent event-level detection. On HAI, the method achieves moderate precision with low false alarm rates and stable event-level performance. On BATADAL, it remains functional without retraining, although precision degrades under domain shift.

The results also show that the LLM does not improve detection performance, but contributes as an interpretation layer. This supports a separation between detection and reasoning, where deterministic DT-based mechanisms provide anomaly signals and LLMs offer complementary explanations.

Overall, the proposed approach provides a lightweight, training-free alternative that prioritizes low false alarm rates and interpretability, making it suitable for early-stage deployment and operational monitoring. Its effectiveness depends on the alignment between constraints and system dynamics, but it retains functionality across datasets without retraining.

Future work includes extending constraint modeling to capture richer interdependencies, improving robustness under heterogeneous conditions, and developing quantitative evaluation protocols for interpretation. Real-time deployment and integration into operational systems remain open directions.

\subsection*{Acknowledgments}
This work is supported by the Research Council of Norway through
the SFI Norwegian Centre for Cybersecurity in Critical Sectors (NORCICS) project no. 310105

\bibliographystyle{unsrtnat}
\bibliography{ref}

@article{alcaraz2024digital,
  title={Digital Twin-assisted anomaly detection for industrial scenarios},
  author={Alcaraz, Cristina and Lopez, Javier},
  journal={International Journal of Critical Infrastructure Protection},
  volume={47},
  pages={100721},
  year={2024},
  doi={10.1016/j.ijcip.2024.100721},
  publisher={Elsevier}
}

@article{birihanu2025explainable,
  title={Explainable correlation-based anomaly detection for Industrial Control Systems},
  author={Birihanu, Ermiyas and Lend{\'a}k, Imre},
  journal={Frontiers in Artificial Intelligence},
  volume={7},
  pages={1508821},
  year={2025},
  doi={https://doi.org/10.3389/frai.2024.1508821},
  publisher={Frontiers Media SA}
}

@article{kausar2025digital,
  title={Digital Twin-Enabled Anomaly Detection for Industrial IoT using Explainable AI},
  author={Kausar, Mohammad ABU},
  journal={Foundation of Computer Science (FCS), NY, USA},
  year={2025},
  doi={10.5120/ijca2025925641}
}

@inproceedings{varghese2022digital,
  title={Digital twin-based intrusion detection for industrial control systems},
  author={Varghese, Seba Anna and Ghadim, Alireza Dehlaghi and Balador, Ali and Alimadadi, Zahra and Papadimitratos, Panos},
  booktitle={2022 IEEE international conference on pervasive computing and communications workshops and other affiliated events (PerCom workshops)},
  pages={611--617},
  year={2022},
  doi={10.1109/PerComWorkshops53856.2022.9767492},
  organization={IEEE}
}

@article{sayghe2025digital,
  title={Digital Twin-Driven Intrusion Detection for Industrial SCADA: A Cyber-Physical Case Study},
  author={Sayghe, Ali},
  journal={Sensors},
  volume={25},
  number={16},
  pages={4963},
  year={2025},
  doi={https://doi.org/10.3390/s25164963},
  publisher={MDPI}
}

@article{krishnaveni2024cyberdefender,
  title={CyberDefender: an integrated intelligent defense framework for digital-twin-based industrial cyber-physical systems},
  author={Krishnaveni, S and Chen, Thomas M and Sathiyanarayanan, Mithileysh and Amutha, B},
  journal={Cluster Computing},
  volume={27},
  number={6},
  pages={7273--7306},
  year={2024},
  doi={https://doi.org/10.1007/s10586-024-04320-x},
  publisher={Springer}
}

@article{farwell2011stuxnet,
  title={Stuxnet and the future of cyber war},
  author={Farwell, James P and Rohozinski, Rafal},
  journal={Survival},
  volume={53},
  number={1},
  pages={23--40},
  year={2011},
  doi={https://doi.org/10.1080/00396338.2011.555586},
  publisher={Taylor \& Francis}
}

@inproceedings{cervini2022don,
  title={Don’t drink the cyber: Extrapolating the possibilities of Oldsmar’s water treatment cyberattack},
  author={Cervini, James and Rubin, Aviel and Watkins, Lanier},
  booktitle={International conference on cyber warfare and security},
  volume={17},
  number={1},
  pages={19--25},
  year={2022},
  doi={10.34190/iccws.17.1.29},
  organization={Academic Conferences International Limited}
}

@article{ayodeji2020new,
  title={A new perspective towards the development of robust data-driven intrusion detection for industrial control systems},
  author={Ayodeji, Abiodun and Liu, Yong-kuo and Chao, Nan and Yang, Li-qun},
  journal={Nuclear engineering and technology},
  volume={52},
  number={12},
  pages={2687--2698},
  year={2020},
  doi={https://doi.org/10.1016/j.net.2020.05.012},
  publisher={Elsevier}
}

@article{bhatt2025enhancing,
  title={Enhancing Anomaly Detection in Industrial Control Systems through Supervised Learning and Explainable Artificial Intelligence.},
  author={Bhatt, Dhruv G and Kyada, Parshad U and Rathore, Rajkumar Singh and Nallakaruppan, MK and Jhaveri, Rutvij H and others},
  journal={Journal of Cybersecurity \& Information Management},
  volume={15},
  number={1},
  doi={10.54216/JCIM.150125},
  year={2025}
}

@article{Kampourakis2025,
  author    = {Kampourakis, Konstantinos E. and Gkioulos, Vasileios and Kavallieratos, Georgios and Lin, Jia-Chun},
  title     = {Digital Twin-Enabled Incident Detection and Response: A Systematic Review of Critical Infrastructures Applications},
  journal   = {International Journal of Information Security},
  year      = {2025},
  volume    = {24},
  number    = {5},
  pages     = {194},
  doi       = {10.1007/s10207-025-01113-0},
  url       = {https://doi.org/10.1007/s10207-025-01113-0}
}

@article{Kampourakis2026,
  author  = {Kampourakis, Konstantinos E. and Gkioulos, Vasileios and Katsikas, Sokratis},
  title   = {Cybersecurity Digital Twins for Industrial Systems: From Literature Synthesis to Framework Design},
  journal = {Information},
  year    = {2026},
  volume  = {17},
  number  = {3},
  pages   = {286},
  doi     = {10.3390/info17030286},
  url     = {https://www.mdpi.com/2078-2489/17/3/286}
}

@article{kampourakis2026systematic,
  title={Systematic Integration of Digital Twins and Constrained LLMs for Interpretable Cyber-Physical Anomaly Detection},
  author={Kampourakis, Konstantinos E and Gkioulos, Vasileios and Katsikas, Sokratis},
  journal={arXiv preprint arXiv:2604.03790},
  doi={https://doi.org/10.48550/arXiv.2604.03790},
  year={2026}
}

@article{krishna2023post,
  title={Post hoc explanations of language models can improve language models},
  author={Krishna, Satyapriya and Ma, Jiaqi and Slack, Dylan and Ghandeharioun, Asma and Singh, Sameer and Lakkaraju, Himabindu},
  journal={Advances in Neural Information Processing Systems},
  volume={36},
  pages={65468--65483},
  doi={10.21203/rs.3.rs-3006112/v1},
  year={2023}
}

@article{majeed2024reliability,
  title={Reliability issues of LLMs: ChatGPT a case study},
  author={Majeed, Abdul and Hwang, Seong Oun},
  journal={IEEE Reliability Magazine},
  volume={1},
  number={4},
  pages={36--46},
  year={2024},
  doi={10.1109/MRL.2024.3420849},
  publisher={IEEE}
}

@inproceedings{shin2020hai,
  title={$\{$HAI$\}$ 1.0:$\{$HIL-based$\}$ augmented $\{$ICS$\}$ security dataset},
  author={Shin, Hyeok-Ki and Lee, Woomyo and Yun, Jeong-Han and Kim, HyoungChun},
  booktitle={13Th USENIX workshop on cyber security experimentation and test (CSET 20)},
  year={2020}
}

@article{taormina2018battle,
  title={Battle of the attack detection algorithms: Disclosing cyber attacks on water distribution networks},
  author={Taormina, Riccardo and Galelli, Stefano and Tippenhauer, Nils Ole and Salomons, Elad and Ostfeld, Avi and Eliades, Demetrios G and Aghashahi, Mohsen and Sundararajan, Raanju and Pourahmadi, Mohsen and Banks, M Katherine and others},
  journal={Journal of Water Resources Planning and Management},
  volume={144},
  number={8},
  pages={04018048},
  year={2018},
  doi={https://doi.org/10.1061/(ASCE)WR.1943-5452.0000969},
  publisher={American Society of Civil Engineers}
}

@article{raman2021hybrid,
  title={A hybrid physics-based data-driven framework for anomaly detection in industrial control systems},
  author={Raman, MR Gauthama and Mathur, Aditya P},
  journal={IEEE Transactions on Systems, Man, and Cybernetics: Systems},
  volume={52},
  number={9},
  pages={6003--6014},
  year={2021},
  doi={10.1109/TSMC.2021.3131662},
  publisher={IEEE}
}

@inproceedings{fung2024attributions,
  title={Attributions for ML-based ICS Anomaly Detection: From Theory to Practice.},
  author={Fung, Clement and Zeng, Eric and Bauer, Lujo},
  booktitle={NDSS},
  doi={10.14722/ndss.2024.23216},
  year={2024}
}

@article{homaei2024review,
  title={A review of digital twins and their application in cybersecurity based on artificial intelligence},
  author={Homaei, Mohammadhossein and Mogoll{\'o}n-Guti{\'e}rrez, {\'O}scar and Sancho, Jos{\'e} Carlos and {\'A}vila, Mar and Caro, Andr{\'e}s},
  journal={Artificial Intelligence Review},
  volume={57},
  number={8},
  pages={201},
  year={2024},
  doi={https://doi.org/10.1007/s10462-024-10805-3},
  publisher={Springer}
}

@article{alhamam2025comprehensive,
  title={A comprehensive review on cybersecurity of digital twins issues, challenges, and future research directions},
  author={Alhamam, Norah and Rahman, MM Hafizur and Aljughaiman, Ahmed},
  journal={IEEE Access},
  volume={13},
  pages={45106--45124},
  year={2025},
  doi={10.1109/ACCESS.2025.3545004},
  publisher={IEEE}
}

@article{mbasso2026digital,
  title={Digital-Twin-Enabled, Time-Aware Anomaly Detection for Industrial Cyber-Physical Systems},
  author={Mbasso, Wulfran Fendzi and Harrison, Ambe and Dagal, Idriss and Jangir, Pradeep and Liu, Zhe and Smerat, Aseel},
  journal={Digital Twins and Applications},
  volume={3},
  number={1},
  pages={e70016},
  year={2026},
  doi={https://doi.org/10.1049/dgt2.70016},
  publisher={Wiley Online Library}
}

@article{lugaresi2023online,
  title={Online validation of digital twins for manufacturing systems},
  author={Lugaresi, Giovanni and Gangemi, Sofia and Gazzoni, Giulia and Matta, Andrea},
  journal={Computers in Industry},
  volume={150},
  pages={103942},
  year={2023},
  doi={https://doi.org/10.1016/j.compind.2023.103942},
  publisher={Elsevier}
}

@article{koay2023machine,
  title={Machine learning in industrial control system (ICS) security: current landscape, opportunities and challenges},
  author={Koay, Abigail MY and Ko, Ryan K L and Hettema, Hinne and Radke, Kenneth},
  journal={Journal of Intelligent Information Systems},
  volume={60},
  number={2},
  pages={377--405},
  year={2023},
  doi={https://doi.org/10.1007/s10844-022-00753-1},
  publisher={Springer}
}

@article{adepoju2022advancing,
  title={Advancing monitoring and alert systems: A proactive approach to improving reliability in complex data ecosystems},
  author={Adepoju, ADEBUSAYO HASSANAT and Austin-Gabriel, BLESSING and Hamza, OLADIMEJI and Collins, ANUOLUWAPO},
  journal={IRE Journals},
  volume={5},
  number={11},
  pages={281--282},
  doi={},
  year={2022}
}

@inproceedings{fung2022perspectives,
  title={Perspectives from a comprehensive evaluation of reconstruction-based anomaly detection in industrial control systems},
  author={Fung, Clement and Srinarasi, Shreya and Lucas, Keane and Phee, Hay Bryan and Bauer, Lujo},
  booktitle={European Symposium on Research in Computer Security},
  pages={493--513},
  year={2022},
  doi={https://doi.org/10.1007/978-3-031-17143-7_24},
  organization={Springer}
}
\end{document}